\documentclass[12pt]{report}
\usepackage[utf8]{inputenc}
\usepackage{blindtext}
\usepackage[letterpaper, total={8.5in, 11in}, margin=2in]{geometry}
\usepackage{ragged2e}
\usepackage{blindtext}
\usepackage{graphicx}
\usepackage{amsmath}
\usepackage{amssymb}
\usepackage{gensymb}
\usepackage{array}
\date{}

\begin{document}
\centering{\huge Downlink Received Power Performance Analysis of IRS for 6G Networks\\
\vspace{24pt}
\large Mobasshir Mahbub, Raed M. Shubair}

\newpage

\RaggedRight{\textbf{\Large 1.\hspace{10pt} Introduction}}\\
\vspace{18pt}
\justifying{\noindent Wireless inventions and approaches, as well as their potential applications, have experienced exponential development and great improvement over the last thirty years. They involve transmitter design and broadcast characteristics [1-18], THz interface and signal enhancement characteristics [19-31], and indoor location techniques and challenges [32-50].

The continuous growth of 5G connection has enabled IoT [51] to connect more things, allowing for larger application areas in a range of sectors. The infrastructure for 5G has been continuously improving [9]. While fifth-generation (5G) [52] as well as following fifth-generation (B5G) [53] techniques are still being deployed internationally, the issues of decreased data rates and restricted capacity have pushed researchers to focus on sixth-generation (6G) solutions [54]-[56]. Developed nations and sophisticated research institutions have begun to investigate and build 6G standards for next-generation network technology. In this scenario, emerging mobile content will function at frequencies greater than 5G technologies. As a result, related network interconnection will be increased, network throughput will increase, and overall consumption of energy should decrease.

The IRS [57] is a flat metasurface composed of a large number of reflective components that has lately attracted academic interest because to its capacity to greatly increase the energy and spectrum efficiency of network communications by changing wireless transmission contexts. IRS pieces with the proper phase shift may convey the incoming wave [58]. IRS generates constructive signal combination as well as detrimental interference suppression at receiving by dynamically modifying reflected signal broadcasts. As a consequence, the receiver's excellence of service (QoS) may be increased.

As a result, in the instance of IoT services and applications, the research compared the performance of standard small cellular connectivity with IRS-enhanced small cellular connectivity in terms of downstream received power.
}
\vspace{18pt}

\RaggedRight{\textbf{\Large 2.\hspace{10pt} Related Literature}}\\
\vspace{18pt}
\justifying{\noindent Xie et al. [59] examined the IRS-assisted MIMO transmitting-based IoT network consisting of numerous low-power IoT devices' weighed sum-rate maximization strategy or challenge. The paper proposed a joint optimization strategy to do the analysis, which decomposes the issue into many sub-problems that may be disposed of alternatively. Yu et al. [60] created an IRS-assisted large network framework with a channel estimation technique. The paper examined the framework's performance. The research discovered that IRS improves end-user performance while expanding network coverage. Mahmoud et al. [61] examined the use of IRS in unmanned airborne vehicle (UAV) powered wireless communications with the goal of improving network coverage and increasing communication consistency and spectral efficiency in IoT. The study specifically developed analytic equations for the stochastic capacity, symbolic rate of error (SER), and likelihood of outages and assessed the formulas in relation to the circumstances of the given network configuration.}

\vspace{18pt}
\RaggedRight{\textbf{\Large 3.\hspace{10pt} Measurement Model}}\\
\vspace{12pt}

\justifying{\noindent Consider a two-tier network with small cell base stations supplying IoT user gadgets (micro cell base station operates beneath a macro cell relying station). In the event of an IRS-assisted system, the aforementioned base stations will service the aforementioned IoT devices that incorporate an IRS intermediate the base platform and the user equipment. $P_t^D$ and $B^D$ are the power used for transmission and bandwidth of the downlink, correspondingly.}

\vspace{12pt}
\RaggedRight{\textit{\large A.\hspace{10pt} Conventional Model}}\\
\vspace{12pt}
\justifying{\noindent The power that is received from a typical small cell for the IoT scenario is represented as follows (Eq. 1) [62]-[64],
}
\vspace{6pt}
\begin{equation}
P_r^{D}= \frac{P_t^{D}\lambda h}{(4\pi)^2 d^\alpha} 
\end{equation}
where $\lambda= c⁄f_c$  is the extension of the wavelength associated with the carrier. $c$ is the speed of light in $ms^{-1}$. $f_c$ represents the carrier resonance in Hz. $h$ is a Rayleigh fading factor following a unit mean distribution with an exponential shape.

$d= \sqrt{(x^{B}-x^{U})^2+(y^{B}-y^{U})^2+(z^{B}-z^{U})^2}$. $\alpha$ is the level of attenuation represented by the exponent. 

\vspace{12pt}
\RaggedRight{\textit{\large B.\hspace{10pt} IRS-Assisted Model}}\\
\vspace{12pt}

\justifying{\noindent In the case of an IRS-assisted small cell for IoT components, the obtained power in the downstream is computed as follows (Eq. 2) [65],}

\begin{equation}
P_r^{D(IRS)} = \frac{P_t^{D(IRS)}G_t G_r G M^2 N^2 d_x d_y \lambda^2 cos(\theta_t) cos(\theta_r) A^2}{64\pi^3(d_1 d_2)^2}
\end{equation}
where $d_1=  \sqrt{(x^{B}-x_i^{I} )^2+(y^{B}-y_i^{I} )^2+(z^{B}-z_i^{I} )^2}$ represents the distance separating the transmitter at $(x^{B},y^{B},z^{B})$ coordinates and IRS at $(x_i^{I},y_i^{I},z_i^{I})$ coordinates. 

$d_2=  \sqrt{(x_i^{IRS}-x_n^{U} )^2+(y_i^{IRS}-y_n^{U} )^2+(z_i^{IRS}-z_n^{U} )^2}$
is the distance separating the IRS to IoT gadgets situated at $(x^{U},y^{U},z^{U})$. $G_t$ is the transmission gain and $G_r$ is the gain of the receiver. $G=\frac{4\pi d_x d_y}{\lambda^2}$  indicates the IRS's dispersion gain. $M$ and $N$ denote the total number of transmit and receive elements. $d_x$ and $d_y$ are IRS elements' length and breadth. $\theta_t$ is the angle of transmission and $\theta_r$  is the receive angle. $A$ is the value of the coefficient of refraction.

\vspace{18pt}

\RaggedRight{\textbf{\Large 4.\hspace{10pt} Numerical Results and Discussions}}\\
\vspace{12pt}
\justifying{\noindent The section of the paper includes the measurement results obtained by MATLAB-based simulation deploying the measurement model and incorporates discussions on the derived results.}

\vspace{12pt}
\RaggedRight{\textit{\large A.\hspace{10pt} Results for Conventional Model}}\\
\vspace{12pt}
\justifying{
Fig. 1 illustrates the measurement of downlink received power throughout the coverage region.}

\vspace{12pt}
\centering{
\includegraphics[height=9.0cm, width=11.5cm]{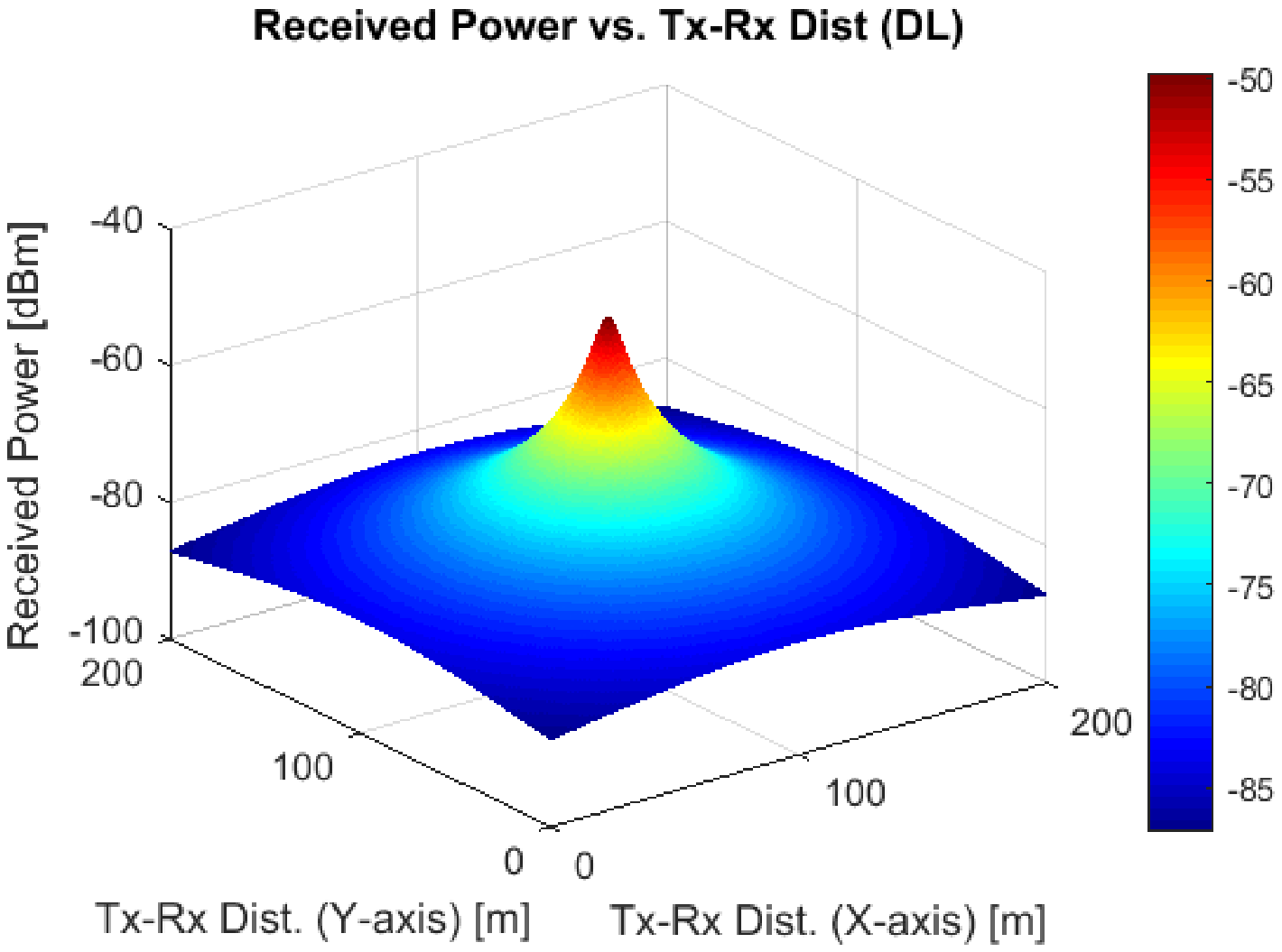}\par

Fig. 3. Transmitter-receiver separation distance vs. received power (non-IRS DL)}

\justifying

\vspace{12pt}
\RaggedRight{\textit{\large B.\hspace{10pt} Results for IRS-Assisted Model}}\\
\vspace{12pt}
\justifying{
Fig. 2 visualizes the downlink received power over the coverage region in terms of varied transmit-receive angles.}
  
\vspace{12pt}
\centering{
\includegraphics[height=9.0cm, width=11.5cm]{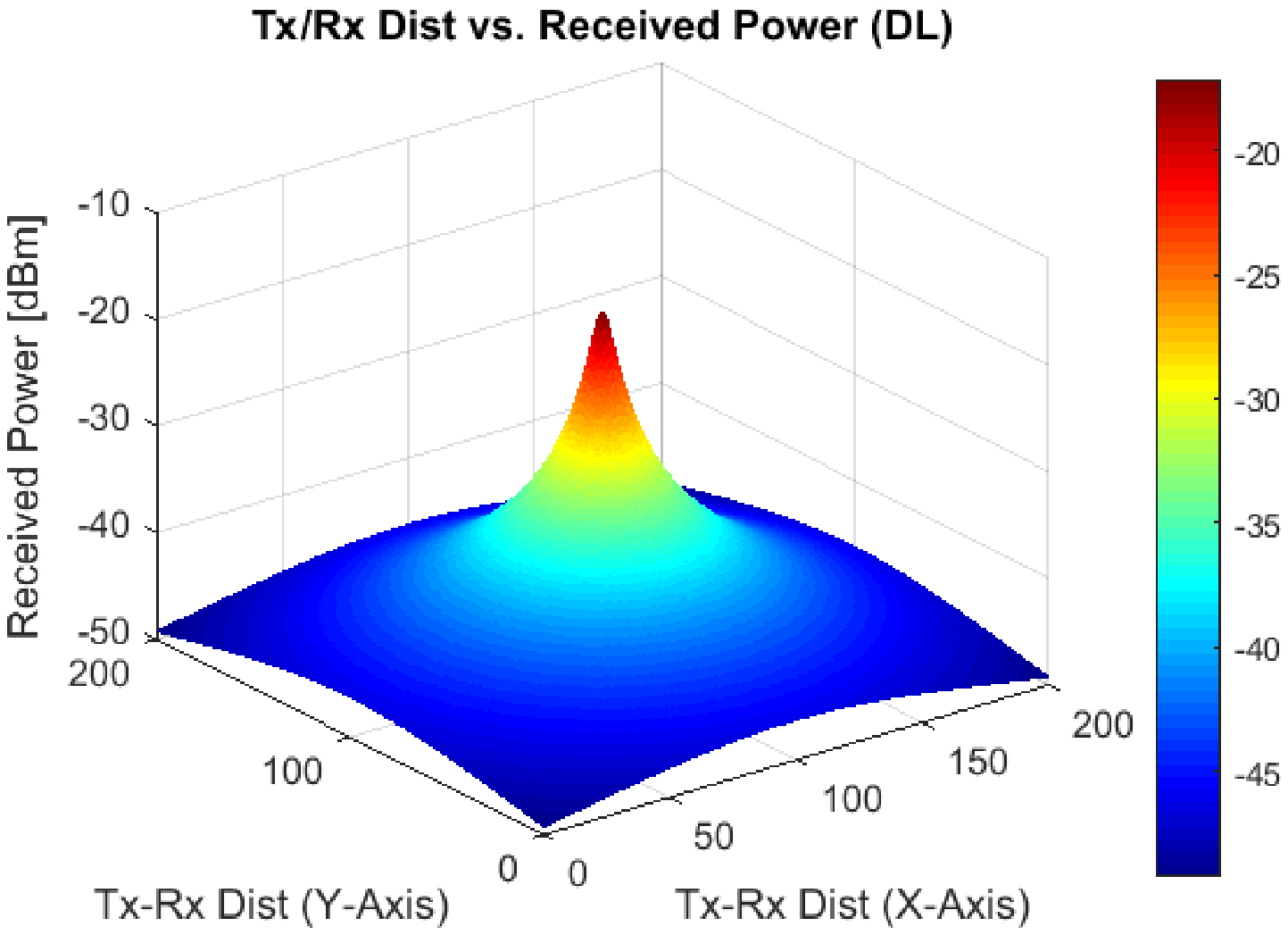}

(a)
}
 
\vspace{12pt}
\centering{
\includegraphics[height=9.0cm, width=11.5cm]{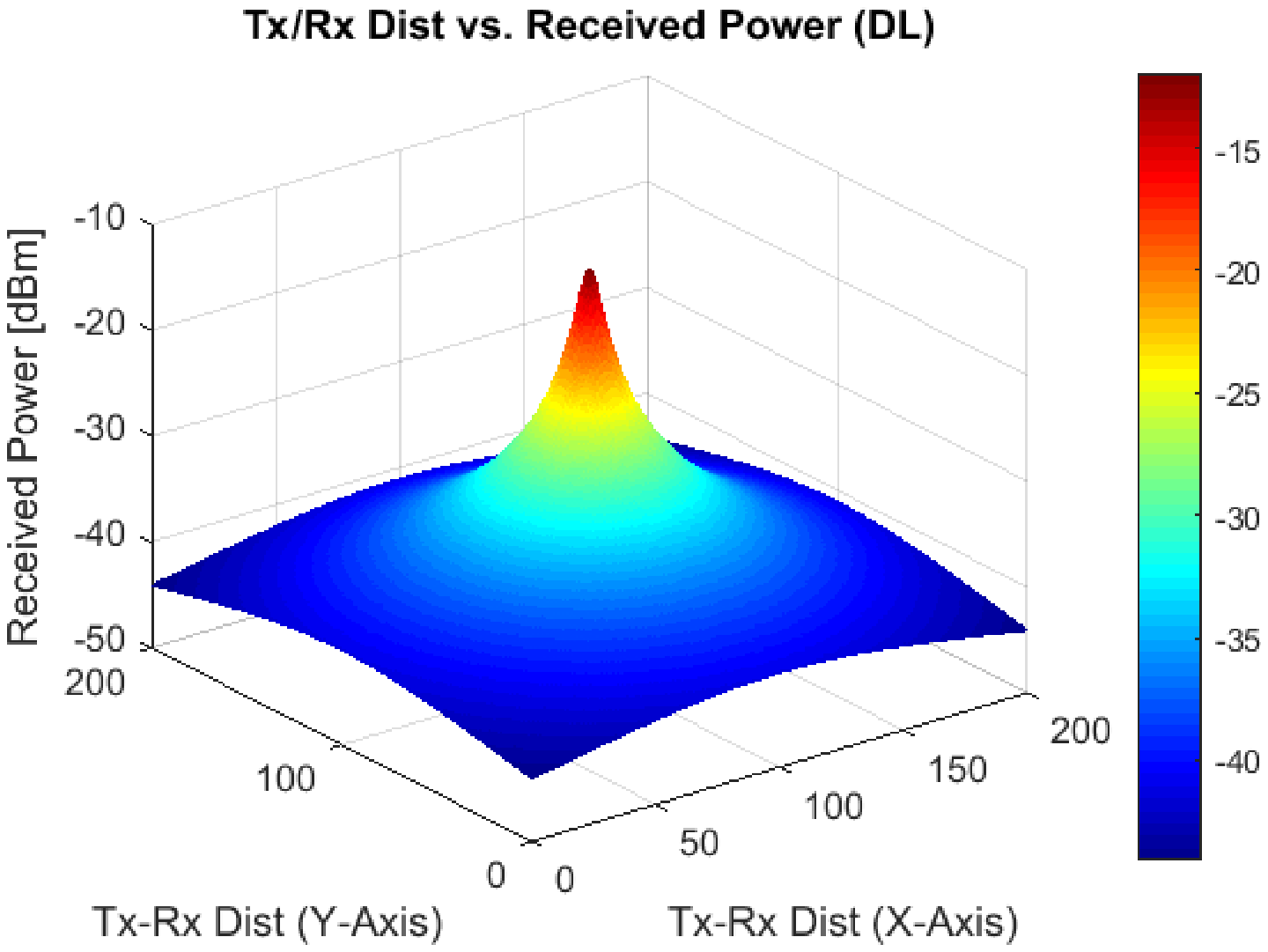}

(b)
}

\vspace{12pt}
\centering{
\includegraphics[height=9.0cm, width=11.5cm]{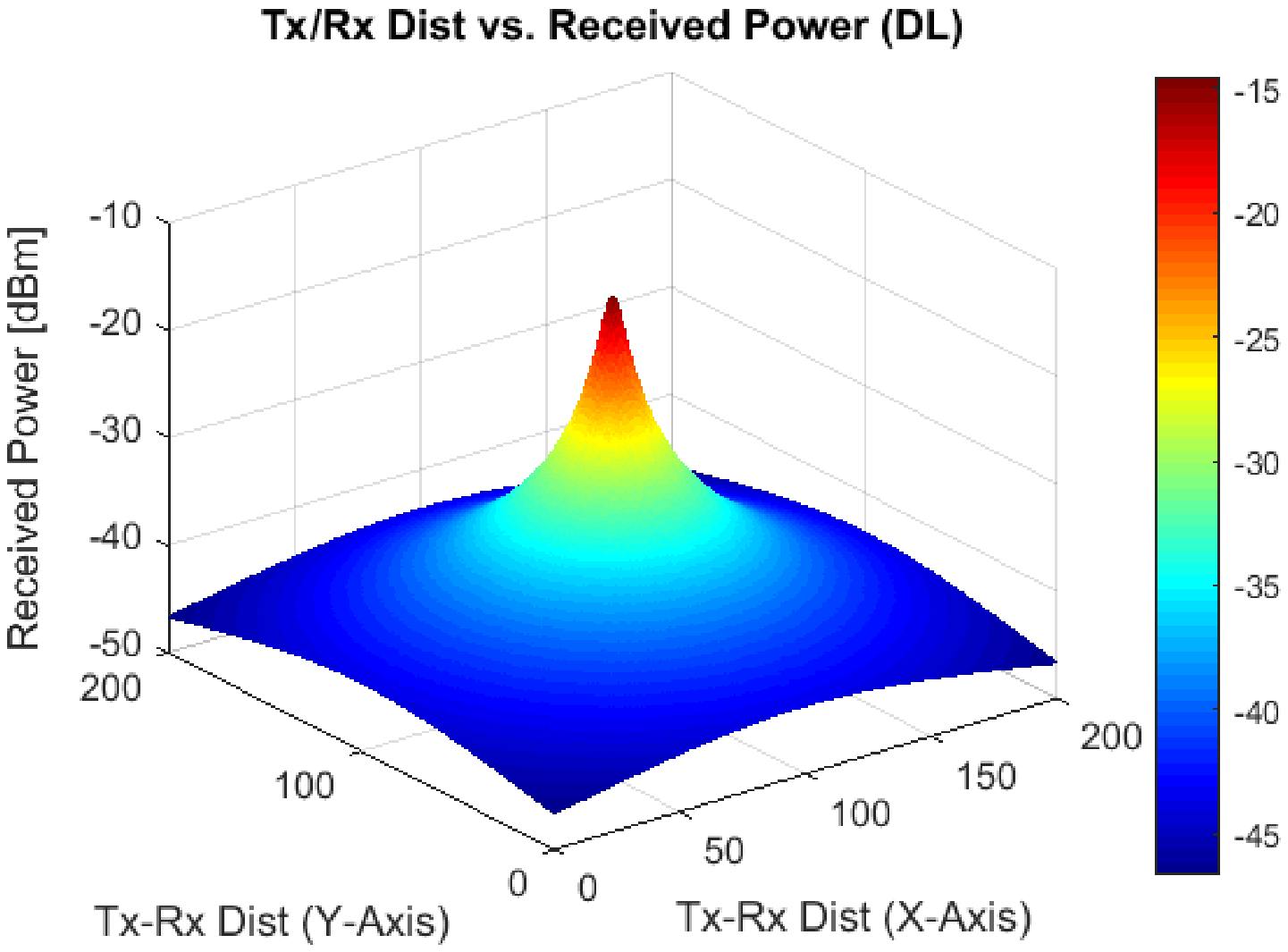}

(c)
}

\justifying{Fig. 13. (a) Transmitter-receiver separation distance vs. received power ($\theta_t= 45\degree$, $\theta_r= 45\degree$), (b) Transmitter-receiver separation distance vs. received power ($\theta_t= 60\degree$, $\theta_r= 60\degree$), (c) Transmitter-receiver separation distance vs. received power ($\theta_t= 45\degree$, $\theta_r= 60\degree$)
The observation of Fig. 2 (a)-(c) states that the $60\degree$ transmit-receive angles (from the base station to IRS and IRS to device) outperform other considered angles in terms of downlink received power. Afterward, $45\degree$ of transmitting and $60\degree$ of receiving angles perform better than other considered transmit-receive angles. In the case of the $60\degree$ transmit-receive angles, the devices located 15 m away from the IRS achieve around -25 dBm of received power, at 50 m the received power in downlink becomes -35 dBm, and at the cell edge, it is -41 to -44 dBm.

In the case of the conventional communication model, the maximum and minimum downlink received powers are -50 and -86 dBm, respectively and in the case of IRS-assisted communication, the maximum and minimum downlink received powers are -12 and -44 dBm, respectively.}

\vspace{18pt}

\RaggedRight{\textbf{\Large 5.\hspace{10pt} Conclusion}}\\
\vspace{12pt}

\justifying{\noindent The research targeted to enhance the coverage for IoT services in the context of forthcoming 6G networks deploying IRS in a micro cell. The work included a literature review of relative existing research to provide insight and figure out research limitations or gaps. It formed a measurement model for both the conventional and IRS-assisted micro cellular network that contains several equations to measure downlink received power.}

\vspace{18pt}

\RaggedRight{\textbf{\Large References}}\\
\justifying{
1.	Mohamed I. AlHajri, Nazar T. Ali, and Raed M. Shubair. "Classification of indoor environments for IoT applications: A machine learning approach." IEEE Antennas and Wireless Propagation Letters 17, no. 12 (2018): 2164-2168.

2.	M. I. AlHajri, A. Goian, M. Darweesh, R. AlMemari, R. M. Shubair, L. Weruaga, and A. R. Kulaib. "Hybrid RSS-DOA technique for enhanced WSN localization in a correlated environment." In 2015 International Conference on Information and Communication Technology Research (ICTRC), pp. 238-241. IEEE, 2015.

3.	Fahad Belhoul, Raed M. Shubair, and Mohammed E. Al-Mualla. "Modelling and performance analysis of DOA estimation in adaptive signal processing arrays." In ICECS, pp. 340-343. 2003.

4.	Ali Hakam, Raed M. Shubair, and Ehab Salahat. "Enhanced DOA estimation algorithms using MVDR and MUSIC." In 2013 International Conference on Current Trends in Information Technology (CTIT), pp. 172-176. IEEE, 2013.

5.	R. M. Shubair and A. Al-Merri. "Robust algorithms for direction finding and adaptive beamforming: performance and optimization." In The 2004 47th Midwest Symposium on Circuits and Systems, 2004. MWSCAS'04., vol. 2, pp. II-II. IEEE, 2004.

6.	Pradeep Kumar Singh, Bharat K. Bhargava, Marcin Paprzycki, Narottam Chand Kaushal, and Wei-Chiang Hong, eds. Handbook of wireless sensor networks: issues and challenges in current Scenario's. Vol. 1132. Berlin/Heidelberg, Germany: Springer, 2020.

7.	R. M. Shubair and W. Jessmi. "Performance analysis of SMI adaptive beamforming arrays for smart antenna systems." In 2005 IEEE Antennas and Propagation Society International Symposium, vol. 1, pp. 311-314. IEEE, 2005.

8.	E. M. Al-Ardi, R. M. Shubair, and M. E. Al-Mualla. "Investigation of high-resolution DOA estimation algorithms for optimal performance of smart antenna systems." (2003): 460-464.

9.	Mohamed AlHajri, Abdulrahman Goian, Muna Darweesh, Rashid AlMemari, Raed Shubair, Luis Weruaga, and Ahmed AlTunaiji. "Accurate and robust localization techniques for wireless sensor networks." arXiv preprint arXiv:1806.05765 (2018).

10.	Goian, Mohamed I. AlHajri, Raed M. Shubair, Luis Weruaga, Ahmed Rashed Kulaib, R. AlMemari, and Muna Darweesh. "Fast detection of coherent signals using pre-conditioned root-MUSIC based on Toeplitz matrix reconstruction." In 2015 IEEE 11th International Conference on Wireless and Mobile Computing, Networking and Communications (WiMob), pp. 168-174. IEEE, 2015.

11.	Zhenghua Chen, Mohamed I. AlHajri, Min Wu, Nazar T. Ali, and Raed M. Shubair. "A novel real-time deep learning approach for indoor localization based on RF environment identification." IEEE Sensors Letters 4, no. 6 (2020): 1-4.

12.	R. M. Shubair, A. Merri, and W. Jessmi. "Improved adaptive beamforming using a hybrid LMS/SMI approach." In Second IFIP International Conference on Wireless and Optical Communications Networks, 2005. WOCN 2005., pp. 603-606. IEEE, 2005.

13.	Satish R. Jondhale, Raed Shubair, Rekha P. Labade, Jaime Lloret, and Pramod R. Gunjal. "Application of supervised learning approach for target localization in wireless sensor network." In Handbook of Wireless Sensor Networks: Issues and Challenges in Current Scenario's, pp. 493-519. Springer, Cham, 2020.

14.	Raed Shubair, and Rashid Nuaimi. "Displaced sensor array for improved signal detection under grazing incidence conditions." Progress in Electromagnetics Research 79 (2008): 427-441.

15.	WafaNjima, Marwa Chafii, ArseniaChorti, Raed M. Shubair, and H. Vincent Poor. "Indoor localization using data augmentation via selective generative adversarial networks." IEEE Access 9 (2021): 98337-98347.

16.	Raed M. Shubair. "Improved smart antenna design using displaced sensor array configuration." Applied Computational Electromagnetics Society Journal 22, no. 1 (2007): 83.

17.	Mohamed Ibrahim Alhajri, N. T. Ali, and R. M. Shubair. "2.4 ghz indoor channel measurements." IEEE Dataport (2018).

18.	Raed M. Shubair, and Hadeel Elayan. "Enhanced WSN localization of moving nodes using a robust hybrid TDOA-PF approach." In 2015 11th International Conference on Innovations in Information Technology (IIT), pp. 122-127. IEEE, 2015.

19.	WafaNjima, Marwa Chafii, and Raed M. Shubair. "Gan based data augmentation for indoor localization using labeled and unlabeled data." In 2021 International Balkan Conference on Communications and Networking (BalkanCom), pp. 36-39. IEEE, 2021.

20.	Mohamed I. AlHajri, Nazar T. Ali, and Raed M. Shubair. "A cascaded machine learning approach for indoor classification and localization using adaptive feature selection." AI for Emerging Verticals: Human-robot computing, sensing and networking (2020): 205.

21.	Hadeel Elayan, Raed M. Shubair, and Asimina Kiourti. "Wireless sensors for medical applications: Current status and future challenges." In 2017 11th European Conference on Antennas and Propagation (EUCAP), pp. 2478-2482. IEEE, 2017.

22.	Raed M. Shubair and Hadeel Elayan. "In vivo wireless body communications: State-of-the-art and future directions." In 2015 Loughborough Antennas \& Propagation Conference (LAPC), pp. 1-5. IEEE, 2015.

23.	Hadeel Elayan, Pedram Johari, Raed M. Shubair, and Josep Miquel Jornet. "Photothermal modeling and analysis of intrabody terahertz nanoscale communication." IEEE transactions on nanobioscience 16, no. 8 (2017): 755-763.

24.	Rui Zhang, Ke Yang, Akram Alomainy, Qammer H. Abbasi, Khalid Qaraqe, and Raed M. Shubair. "Modelling of the terahertz communication channel for in-vivo nano-networks in the presence of noise." In 2016 16th Mediterranean Microwave Symposium (MMS), pp. 1-4. IEEE, 2016.

25.	Hadeel Elayan, Raed M. Shubair, and Asimina Kiourti. "On graphene-based THz plasmonic nano-antennas." In 2016 16th mediterranean microwave symposium (MMS), pp. 1-3. IEEE, 2016.

26.	Hadeel Elayan, Cesare Stefanini, Raed M. Shubair, and Josep Miquel Jornet. "End-to-end noise model for intra-body terahertz nanoscale communication." IEEE transactions on nanobioscience 17, no. 4 (2018): 464-473.

27.	Hadeel Elayan, and Raed M. Shubair. "On channel characterization in human body communication for medical monitoring systems." In 2016 17th International Symposium on Antenna Technology and Applied Electromagnetics (ANTEM), pp. 1-2. IEEE, 2016.

28.	Hadeel Elayan, Raed M. Shubair, and Nawaf Almoosa. "In vivo communication in wireless body area networks." In Information Innovation Technology in Smart Cities, pp. 273-287. Springer, Singapore, 2018.

29.	Mayar Lotfy, Raed M. Shubair, Nassir Navab, and Shadi Albarqouni. "Investigation of focal loss in deep learning models for femur fractures classification." In 2019 International Conference on Electrical and Computing Technologies and Applications (ICECTA), pp. 1-4. IEEE, 2019.

30.	S. Elmeadawy, and R. M. Shubair. "Enabling technologies for 6G future wireless communications: Opportunities and challenges. arXiv 2020." arXiv preprint arXiv:2002.06068.

31.	Abdul Karim Gizzini, Marwa Chafii, Ahmad Nimr, Raed M. Shubair, and Gerhard Fettweis. "Cnn aided weighted interpolation for channel estimation in vehicular communications." IEEE Transactions on Vehicular Technology 70, no. 12 (2021): 12796-12811.

32.	Nishtha Chopra, Mike Phipott, Akram Alomainy, Qammer H. Abbasi, Khalid Qaraqe, and Raed M. Shubair. "THz time domain characterization of human skin tissue for nano-electromagnetic communication." In 2016 16th Mediterranean Microwave Symposium (MMS), pp. 1-3. IEEE, 2016.

33.	Abdul Karim Gizzini, Marwa Chafii, Shahab Ehsanfar, and Raed M. Shubair. "Temporal Averaging LSTM-based Channel Estimation Scheme for IEEE 802.11 p Standard." arXiv preprint arXiv:2106.04829 (2021).

34.	M. Saeed Khan, A-D. Capobianco, Sajid M. Asif, Adnan Iftikhar, Benjamin D. Braaten, and Raed M. Shubair. "A pattern reconfigurable printed patch antenna." In 2016 IEEE International Symposium on Antennas and Propagation (APSURSI), pp. 2149-2150. IEEE, 2016.

35.	M. Saeeed Khan, Adnan Iftikhar, Sajid M. Asif, Antonio‐Daniele Capobianco, and Benjamin D. Braaten. "A compact four elements UWB MIMO antenna with on‐demand WLAN rejection." Microwave and Optical Technology Letters 58, no. 2 (2016): 270-276.

36.	Muhammad Saeed Khan, Adnan Iftikhar, Raed M. Shubair, Antonio-D. Capobianco, Benjamin D. Braaten, and Dimitris E. Anagnostou. "Eight-element compact UWB-MIMO/diversity antenna with WLAN band rejection for 3G/4G/5G communications." IEEE Open Journal of Antennas and Propagation 1 (2020): 196-206.

37.	Amjad Omar, and Raed Shubair. "UWB coplanar waveguide-fed-coplanar strips spiral antenna." In 2016 10th European Conference on Antennas and Propagation (EuCAP), pp. 1-2. IEEE, 2016.

38.	Ala Eldin Omer, George Shaker, Safieddin Safavi-Naeini, Georges Alquié, Frédérique Deshours, Hamid Kokabi, and Raed M. Shubair. "Non-invasive real-time monitoring of glucose level using novel microwave biosensor based on triple-pole CSRR." IEEE Transactions on Biomedical Circuits and Systems 14, no. 6 (2020): 1407-1420.

39.	Muhammad S. Khan, Syed A. Naqvi, Adnan Iftikhar, Sajid M. Asif, Adnan Fida, and Raed M. Shubair. "A WLAN band‐notched compact four element UWB MIMO antenna." International Journal of RF and Microwave Computer‐Aided Engineering 30, no. 9 (2020): e22282.

40.	R. Karli, H. Ammor, R. M. Shubair, M. I. AlHajri, and A. Hakam. "Miniature Planar Ultra-Wide-Band Microstrip Patch Antenna for Breast Cancer Detection." Skin 1 (2016): 39.

41.	Mohammed S Al-Basheir, Raed M Shubai, and Sami M. Sharif. "Measurements and analysis for signal attenuation through date palm trees at 2.1 GHz frequency." (2006).

42.	Amjad Omar, Maram Rashad, Maryam Al-Mulla, Hussain Attia, Shaimaa Naser, Nihad Dib, and Raed M. Shubair. "Compact design of UWB CPW-fed-patch antenna using the superformula." In 2016 5th International Conference on Electronic Devices, Systems and Applications (ICEDSA), pp. 1-4. IEEE, 2016.

43.	Muhammad S. Khan, Adnan Iftikhar, Raed M. Shubair, Antonio D. Capobianco, Benjamin D. Braaten, and Dimitris E. Anagnostou. "A four element, planar, compact UWB MIMO antenna with WLAN band rejection capabilities." Microwave and Optical Technology Letters 62, no. 10 (2020): 3124-3131.

44.	Omar Masood Khan, Qamar Ul Islam, Raed M. Shubair, and Asimina Kiourti. "Novel multiband Flamenco fractal antenna for wearable WBAN off-body communication applications." In 2018 International Applied Computational Electromagnetics Society Symposium (ACES), pp. 1-2. IEEE, 2018.

45.	Raed M. Shubair, Amer Salah, and Alaa K. Abbas. "Novel implantable miniaturized circular microstrip antenna for biomedical telemetry." In 2015 IEEE International Symposium on Antennas and Propagation \& USNC/URSI National Radio Science Meeting, pp. 947-948. IEEE, 2015.

46.	Yazan Al-Alem, Ahmed A. Kishk, and Raed M. Shubair. "One-to-two wireless interchip communication link." IEEE Antennas and Wireless Propagation Letters 18, no. 11 (2019): 2375-2378.

47.	Yazan Al-Alem, Raed M. Shubair, and Ahmed Kishk. "Efficient on-chip antenna design based on symmetrical layers for multipath interference cancellation." In 2016 16th Mediterranean Microwave Symposium (MMS), pp. 1-3. IEEE, 2016.

48.	Yazan Al-Alem, Raed M. Shubair, and Ahmed Kishk. "Clock jitter correction circuit for high speed clock signals using delay units a nd time selection window." In 2016 16th Mediterranean Microwave Symposium (MMS), pp. 1-3. IEEE, 2016.

49.	Melissa Eugenia Diago-Mosquera, Alejandro Aragón-Zavala, Fidel Alejandro Rodríguez-Corbo, Mikel Celaya-Echarri, Raed M. Shubair, and Leyre Azpilicueta. "Tuning Selection Impact on Kriging-Aided In-Building Path Loss Modeling." IEEE Antennas and Wireless Propagation Letters 21, no. 1 (2021): 84-88.

50.	Yazan Al-Alem, Yazan, Ahmed A. Kishk, and Raed M. Shubair. "Employing EBG in Wireless Inter-chip Communication Links: Design and Performance." In 2020 IEEE International Symposium on Antennas and Propagation and North American Radio Science Meeting, pp. 1303-1304. IEEE, 2020.

51.	S. Li et al., “5G Internet of Things: A survey,” Journal of Industrial Information Integration, vol. 10, pp. 1-9, June 2018.

52.	F. Al-Ogaili and R. M. Shubair, "Millimeter-wave mobile communications for 5G: Challenges and opportunities," 2016 IEEE International Symposium on Antennas and Propagation (APSURSI), 2016, pp. 1003-1004.

53.	C. R. Storck and F. Duarte-Figueiredo, "A Survey of 5G Technology Evolution, Standards, and Infrastructure Associated With Vehicle-to-Everything Communications by Internet of Vehicles," in IEEE Access, vol. 8, pp. 117593-117614, 2020.

54.	A. Dogra, R. K. Jha and S. Jain, "A Survey on Beyond 5G Network With the Advent of 6G: Architecture and Emerging Technologies," in IEEE Access, vol. 9, pp. 67512-67547, 2021.

55.	C. D. Alwis et al., "Survey on 6G Frontiers: Trends, Applications, Requirements, Technologies and Future Research," in IEEE Open Journal of the Communications Society, vol. 2, pp. 836-886, 2021.

56.	M. Alsabah et al., "6G Wireless Communications Networks: A Comprehensive Survey," in IEEE Access, vol. 9, pp. 148191-148243, 2021.
57.	S. Zhang and R. Zhang, "Intelligent Reflecting Surface Aided Multi-User Communication: Capacity Region and Deployment Strategy," in IEEE Transactions on Communications, vol. 69, no. 9, pp. 5790-5806, Sept. 2021.

58.	C. You, B. Zheng and R. Zhang, "Channel Estimation and Passive Beamforming for Intelligent Reflecting Surface: Discrete Phase Shift and Progressive Refinement," in IEEE Journal on Selected Areas in Communications, vol. 38, no. 11, pp. 2604-2620, Nov. 2020.

59.	X. Xie et al., "A Joint Optimization Framework for IRS-assisted Energy Self-sustainable IoT Networks," in IEEE Internet of Things Journal.

60.	G. Yu, X. Chen, C. Zhong, H. Lin and Z. Zhang, "Large Intelligent Reflecting Surface Enhanced Massive Access for B5G Cellular Internet of Things," 2020 IEEE 91st Vehicular Technology Conference (VTC2020-Spring), 2020, pp. 1-5.

61.	A. Mahmoud, S. Muhaidat, P. C. Sofotasios, I. Abualhaol, O. A. Dobre and H. Yanikomeroglu, "Intelligent Reflecting Surfaces Assisted UAV Communications for IoT Networks: Performance Analysis," in IEEE Transactions on Green Communications and Networking, vol. 5, no. 3, pp. 1029-1040, Sept. 2021.

62.	T. Mir, L. Dai, Y. Yang, W. Shen and B. Wang, "Optimal FemtoCell Density for Maximizing Throughput in 5G Heterogeneous Networks under Outage Constraints," 2017 IEEE 86th Vehicular Technology Conference (VTC-Fall), Toronto, ON, Canada, 2017, pp. 1-5.

63.	N. Hassan and X. Fernando, "Interference Mitigation and Dynamic User Association for Load Balancing in Heterogeneous Networks," in IEEE Transactions on Vehicular Technology, vol. 68, no. 8, pp. 7578-7592, Aug. 2019.

64.	M. Mozaffari, W. Saad, M. Bennis and M. Debbah, "Optimal Transport Theory for Cell Association in UAV-Enabled Cellular Networks," in IEEE Communications Letters, vol. 21, no. 9, pp. 2053-2056, Sept. 2017.

65.	W. Tang et al., "Wireless Communications With Reconfigurable Intelligent Surface: Path Loss Modeling and Experimental Measurement," in IEEE Transactions on Wireless Communications, vol. 20, no. 1, pp. 421-439, Jan. 2021.
}

\end{document}